\input amstex
\documentstyle{amsppt}{\catcode`@=11\gdef\logo@{}}
\NoRunningHeads
\TagsOnRight
\NoBlackBoxes
\define\vv{\vskip4mm}
\document
\topmatter
\title The Expansive Nondecelerative Universe - gravitational effects and their
manifestation in black holes evaporation and far-infrared
spectra\endtitle

\author {Jozef \v S{}ima, and Miroslav S\'{u}ken\'{\i}k}\endauthor

\address{Slovak Technical University,\newline
 Faculty of Chemical Technology\newline
 Radlinsk\'{e}ho 9\newline
 81237 Bratislava,\newline Slovak Republic}\endaddress
\email{sima\@chelin.chtf.stuba.sk}\endemail
\abstract
The paper summarizes the background of Expensive Nondecelerative
Universe model and its main consequences for gravitation. Applying the Vaidya
metrics, the model allows for the localization and determination of the
density and quantity of gravitational energy created by a body with the mass m
in the distance r. The consequences are manifested both in a macrosystem
(Hawking's phenomenon of black holes evaporation) and microworld phenomenon
(far-infrared spectral properties)
\endabstract
\endtopmatter

\subhead 1. Introduction\endsubhead
\vv

Questions of the Universe creation, evolution and future have formed the
central part of mankind interest from the beginning of appearance of human
beings on the Earth. Naturally, a number of different approaches not
contradicting to generally accepted natural laws has emerged. One of the
models, offering an explanation of the mechanism of matter creation in the
Universe [Skalsk\'{y}, S\'{u}ken\'{\i}k, 1991b)], allowing to precise the
exact values of fundamental physical constants and present parameters of the
Universe [Skalsk\'{y}, S\'{u}ken\'{\i}k, 1992a,b,d, 1993a, 1994, 1996] and
hypothesing its future developmentis our model [Skalsk\'{y}, S\'{u}ken\'{\i}k,
1992c, 1993b] of Expansive Nondecelerate Universe (hereafter referred to as
ENU). In the present contribution, a general overview of the model is given, a
possibility to obtain the famous Hawking's relation concerning the black holes
evaporation [Hawking, 1980, 1988] in an independent way is documented, and the
issue of new spectral bands in low-temperature far-infrared spectra is discussed.
\vv

\subhead 2. Background of the ENU model\endsubhead
\vv

Solving Einstein's equations of field (1915) by means of the Robertson-Walker
metrics, Friedmann (1922, 1924) obtained equations of universe dynamics. In
the ENU model it is stated and rationalized [Skalsk\'{y}, S\'{u}ken\'{\i}k,
1991a] that the Universe, throughout the whole expansive evolutionary phase,
expands by the velocity of light, and the gauge factor a can be thus expressed as

$$a = c.t \tag"(1)"$$

where t is the cosmological time. In this approach, for cases with the
curvature index k and cosmologic member being of zero value
$$k = 0 \tag"(2)"$$
$$\Lambda= 0\tag"(3)"$$
Friedmann's equations describing ENU dynamics
$$(\dot a)^{2} =\frac{8}{3} \pi G\rho a^{2} -kc^{2} +\frac{1}{3} \Lambda^{} a^{2}
c^{2}\tag"(4)" $$
$$2a.\ddot a+(\dot a)^{2} =-\frac{8\pi Gpa^{2} }{c^{2} } -kc^{2} +\Lambda a^{2} c^{2}
\tag"(5)"$$
can be rewritten [Skalsk\'{y}, S\'{u}ken\'{\i}k, 1991a] as follows:
$$c^{2} =\frac{8\pi G\rho a^{2} }{3} =-\frac{8\pi Gpa^{2} }{c^{2} }
\tag"(6)"$$
In the above equations, is the mean (critical) mass density of the Universe
(at present, $\rho \thickapprox 10^{-26}$ kg m$^{-3}$) p is the pressure. Since the energy density is
$$\epsilon=\rho c^2 \tag"(7)"$$
in accordance with the General Theory of Relativity and (6), the Universe with
total zero and local non-zero energy has to be described by the state equation
[Skalsk\'{y}, S\'{u}ken\'{\i}k, 1993b]

$$\epsilon+ 3p = 0 \tag"(8)"$$

Providing that the velocity of the Universe expansion is equal to c, the total
gravitational force must be equal to zero (compare 1 and 8). Equation (6)
leads then to the expression for
$$\rho=\frac{3c^{2} }{8\pi Ga^{2} } \tag"(9)"$$
and, due to (2), for representing the density of the Universe with the mass
$M_u$, it must be valid at the same time
$$\rho=\frac{3M_{u} }{4\pi a^{3} } \tag"(10)"$$
Taking into account the expressions (9) and (10), for the gravitational radius
it then holds
$$a=ct=\frac{2GM_{u} }{c^{2} }  \tag"(11)"$$
Since a is increasing in time (at present, a $\thickapprox 1.3\times 10^{26} m$),$ M_u$ (its present
value approaches $8.6\times 10^{52}$ kg) must increase as well, i.e. in ENU, the creation
of matter occurs [Skalsk\'{y}, S\'{u}ken\'{\i}k, 1993b]. The total energy of
the Universe must, however, be exactly zero [Hawking, 1980, 1988]. It is
achieved by a simultaneous gravitational field creation, the energy of which
is $E < 0$. The fundamental conservation law is thus observed.

For weak gravitational fields, the density of gravitational energy is in first
approximation defined by Tolman's equation:
$$\epsilon_{g} =-\frac{Rc^{4} }{8\pi G} \tag"(12)"$$
where R is the scalar curvature. In Schwarzschild metrics, R = 0, i.e.
gravitational energy is not localizable outside a body since
$\epsilon_g = 0$. Due to the
matter creation, Schwarzschild metrics must be replaced by Vaidya metrics
[Vaidya, 1951, Shipov, 1993], applying of which to ENU leads to relation:
$$R=\frac{6G\dot m_{t} }{r^{2} c^{3} } =\frac{6Gm}{tr^{2} c^{3} } =\frac{3r_{g(m)}
}{ar^{2} }  \tag"(13)"$$
where R is the scalar curvature in the distance r for a body having the mass
m; rg(m) is the gravitational radius of the given body.

Substituting R in (12) for the last term of (13), the density of energy
$\epsilon_g$
induced by such a body in the distance $r$ can be expressed as:
$$\epsilon_{g} =-\frac{3mc^{2} }{4\pi ar^{2} }  \tag"(14)"$$
Within the limits of ENU model it is thus possible to localize and determine
$\epsilon_g$.
For an energy quantum $E_g$ with the density $\epsilon_g$ it holds:

$$\epsilon_{g} =\frac{3E_{g} }{4\pi\lambda^{3} }  \tag"(15)"$$
where the Compton's wave may be expressed as:
$$\lambda=\frac{hc}{2\pi E_{g} }  \tag"(16)"$$
Substituting in (15) for (16) and comparing the result with (14), the
expression for an energy quantum is obtained:
$$\left|  \right.  E_{g} \left.  \right|  =\int\epsilon_{g} .dV =\left(
\frac{mh^{3} c^{5} }{8\pi^{3} ar^{2} } \right)  ^{1/4} \tag"(17)"$$
where $E_g$ is the quantum of gravitational energy created by a body with the
mass $ m$ in the distance $r$. Relation (17) is in conformity with the limiting
values: the maximum energy is represented by the Planck energy, the minimum
energy equals the energy of a photon with the wavelength identical to the
Universe dimension $(a =\lambda )$.
\vv

\subhead 3. Black Holes Evaporation\endsubhead
\vv

Based on thermodynamics and quantum mechanics, Hawking (1980, 1988) found that
a black hole with the radius $r_{BH}$ evaporates via emitting the photons with the energy

$$E=\frac{hc}{2\pi r_{BH} } \tag"(18)"$$
Such an evaporation bears the name Hawking's phenomenon. The total energy
output P per second can be expressed as:
$$P=\frac{hc^{2} }{2\pi r_{BH}^{2} } \tag"(19)"$$
According to Hawking, a black hole with the mass $m_{BH}$ and the gravitational
radius $r_{BH}$ would fully evaporate in time:
$$t=\frac{8\pi G^{2} m_{BH}^{3} }{hc^{4} } \tag"(20)"$$
In order not to violate the second law of thermodynamics, (entropy of a system
involving black hole cannot decrease during any of its evolution phase
[Bokenstein, 1980]) application of (17) to black holes must be written as follows:

$$\left(  \frac{m_{BH} h^{3} c^{5} }{8\pi^{3} ar_{BH}^{2} } \right)  ^{1/4}
\geq\frac{hc}{2\pi r_{BH} } \tag"(21)"$$
When no entropy change would occur (a limiting case), the symbol = is applied
and relation (21) is simplified to
$$\frac{m_{BH} c}{a} =\frac{h}{2\pi r_{BH} ^{2} } \tag"(22)"$$
Taking into account that $t = a/c$ (1) and expressing $r_{BH}$ from (13) as
$$r_{BH} =\frac{2G.m_{BH} }{c^{2} } \tag"(23)"$$
the original Hawking's equation (20) can be derived.

Given that for real processes there must be the sign $>$ in (21), our model
offers the following rationalization of the issue:

i) Hawking's relation (20) correctly predicts the time for a black hole
evaporation in cases when no overall entropy change occurs within the entire
period of the evaporation.

ii) As a consequence of overall entropy increase in real processes, no total
evaporation of a black hole can actually occur since the total energy of
photons released by a black hole within its evaporation, represented by the
right side of (21), is lower than the total gravitational energy created
within the process, represented by the left side of (21). This is in
accordance with the fact that no total evaporation of black holes has been
experimentally observed.

iii) In a given cosmologic time ($t\thickapprox 4.3\times
10^{17}$s) only the black holes with mBH
higher than the minimum mass $m_{BH}\thickapprox 2.7\times
10^{12}$ kg (estimated at the condition that
the both sides of (21) are of equal values) can exist. Our model thus permits
to determine the limiting value of the mass of real black holes in real
cosmologic time.
\vv

\subhead 4. Low-temperature far-infrared spectra\endsubhead
\vv

The equation (17) is of a general application both in macroworld and
microworld. The validity of equation (17) in the microworld can be checked by
means of far-infrared or Raman spectroscopy. For this purpose we substitute
the mass m in equation (17) by a mass of the corresponding atomic nucleus.
Relating the mass m via proton mass $m_p$ and mass number A
$$m = m_p.A\tag"(24)"$$
and expressing further the radius $r$ of a nucleus as
$$r = r_o.A^{1/3}\tag"(25)"$$
where $r_0$ is the Compton length of -mesone $(r_0\thickapprox 1.4\times
10^{-15} m)$
we can arrange the relation (17) into the form
$$\vert E_g\vert =\frac{h.\omega}{2\pi} =\frac{(m_{p} .h^{3} .c^{5} )^{1/4} }{(8\pi^{3}
.a.r_{o} ^{2} )^{1/4} } .A^{1/12} \tag"(26)"$$
In some experimental techniques (e.g. infrared and Raman spectroscopy) the
energy is expressed as wavenumbers in $cm^-1$ and for such cases the equation
(26) can be written as

$$v= \frac{1}{200\pi} .\frac{(2\pi.m_{p} .c)^{1/4} }{(a.r_{o} ^{2} .h)^{1/4} }
.A^{1/12} \tag"(27)"$$
Introducing numerical values for the constant parameters into (27) for (in
$cm^{-1}$) we get
$$v\thickapprox 105.A^{1/12}\tag"(28)"$$

It follows from the above equation (28) that for atoms of naturally occurring
elements with the mass number from A = 1 (hydrogen $_1$H atom) to 238 (uranium
$_{238}$U) the energy due to their gravitation should span in 105
$cm^{-1} - 165 cm^{-1}$,
i.e. in the domain of far-infrared and Raman spectroscopies. It should be
pointed out that contrary to the changes in vibrational and rotational
energies, no theoretical background dealing with gravitation and the mentioned
spectrospopic methods has been elaborated so far. Our attempt to use these
methods to detect gravitational field is based on the following postulates and
assumptions (which are still waiting for theoretical elaboration) :

a) the incident radiation will interact with gravitational field created by
atomic nuclei and consequences of such interaction will manifest in the spectra,

b) the interaction will be observable mainly at low temperatures when the
presence of ``hot bands`` and peaks of lattice vibrations will be suppressed,

c) due to the effects such as couplings with other motions, low transition
probability and different bonding of atoms in investigated compounds, some
peaks may be hidden, undetectable or shifted if comparing to the calculated value.

Searching the literature devoted to far-infrared and Raman spectroscopy we
found several papers presenting spectra scanned at various temperatures in the
above mentioned region. In the spectra, the new peaks, which are not
attributable to the vibrations predictable by the group theory and normal mode
analysis, emerged at low temperatures. This phenomenon is documented by four examples.

The first one is the Raman spectrum of Hg$_2$(NO$_3$)$_2$.2D$_2$O measured at 295 K and 12
K (Cheetham and Day, 1991). It is obvious that new peaks positioned at 111
$cm^{-1}$ and 163 $cm^{-1}$ appeared at 12 K that is in excellent agreement with the
calculated values for deuterium and mercury, respectively. A peak at 138
cm$^{-1}$
can be assigned to the nitrogen and oxygen atoms. A small shift versus the
calculated values (131 cm$^{-1}$, 132 cm$^{-1}$) might be a consequence of a coupling or
a fact that the bond order N-O is higher than 1.

Temperature dependence of the Raman spectra of 1,4-cyclohexadiene (Hagemann et
al., 1985) exhibits at 7 K the presence of a new peak located at 105 cm$^{-1}$ and
a shoulder at 130 cm$^{-1}$ that, in accordance with the values calculated using
equation (28), can be due to hydrogen and carbon, respectively.

As a further example (Futamata et al., 1983) the spectra of alkali metal salts
of tetracyanoquinodimethanide can be introduced. It is obvious that decreasing
the temperature from 298 K to 30 K gives rise to a formation of new spectral
peaks, localized at 124 cm$^{-1}$, 137 cm$^{-1}$ and 144 cm$^{-1}$ for
Li$^+$, Na+ and K$^+$ salt,
respectively. It is worth mentioning that the peak of sodium compound is
absent in the spectra scanned at higher temperatures.

At the end, a structured peak centered at 156 cm$^{-1}$ in the Raman spectra of
three compounds containing the complex anion [Au(CN)$_2$]$^-$ can be mentioned. The
peak was assigned (Assefa, 1994) to a lattice vibration. The fact that its
position does not change with a counter-cation and, moreover, is very near to
the calculated position (162 cm$^{-1}$) suggests that the gravitational effect
might contribute to the peak. The issue of low-temperature far-infrared
spectra can be concluded as follows:

i) The paper documents the potentials of spectroscopic methods to detect
energy changes associated with gravitation, i.e. experimentally verify the
theoretical background elaborated and published in our previous papers. To
accept a more definite conclusion relating to this field, a thorough searching
in the literature must be performed and the obtained data evaluated. It should
be pointed out that the data should meet at least the following requirements:
the spectra should be scanned in the region 100 - 200 cm$^{-1}$ at various
temperatures reaching down to 30 K and lower in order to identify new peaks in
the region; the spectra should be interpreted in more detail in order to
eliminate peaks attributable to normal modes. Papers with the spectra meeting
the above requirements are, however, very rare.

ii) The results obtained so far are of preliminary nature, however, they seem
to be promising enough to stimulate further theoretical and experimental
research in the field.

iii) Based on the derived relation (28), peaks due to the presence of all the
elements can be predicted, e.g., at 134 cm$^{-1}$, 141 cm$^{-1}$, 151 cm$^{-1}$ and 157
cm$^{-1}$
for single bonded fluorine, chlorine, bromine and iodine, respectively.

\vv
\subhead 5. Conclusions\endsubhead
\vv

The development of theoretical elaboration and rationalization of the issues
concerning gravitation, having been started with classic pioneering works at
the two first decades of our century, significantly decelerated. One of the
reasons might be a unsuccess in many direct experimental observations of
gravitational effects, the other can lie in sticking to metrics which do not
allow to deal with the phenomena involving matter creation.

In this paper we show that a usage of Vaidya (or more generally, Vaidya-like)
metrics removes some of the obstacles, unables to localize and quantify
gravitational energy and allows to do further steps both in the theory of
gravitation and the theoretical rationalization of experimentall observed
phenomena. These new possibilities are documented both in the macroworld
(black holes evaporation) and in the microworld (far-infrared spectra)
systems. The presented results are of preliminary nature and we hope they will
stimulate a development of new theoretical approaches and evaluation of
experimental data.

\enddocument